\pdfoutput=1

\documentclass[11pt]{article}

\usepackage[]{acl2023}

\usepackage{times}
\usepackage{latexsym}

\usepackage[utf8]{inputenc}

\usepackage{microtype}

\usepackage{subfigure}
\usepackage{graphicx}
\usepackage{booktabs} 
\usepackage{multicol}
\usepackage{multirow}
\usepackage{pbox}
\usepackage{subfigure}
\usepackage{amsmath}
\usepackage{tabularx}
\usepackage{array}
\usepackage{cellspace}
\usepackage{amsfonts}
\usepackage{amssymb}
\usepackage{algpseudocode}
\usepackage{tabularray}
\usepackage[export]{adjustbox}
\usepackage{soul}
\usepackage{color}
\usepackage[ruled,linesnumbered]{algorithm2e}
\usepackage[T1]{fontenc}
\usepackage{courier}
\usepackage{mathtools}
\usepackage{enumitem}
\usepackage{titling}

\newcommand{\EndocIcon}{\includegraphics[width=1em]{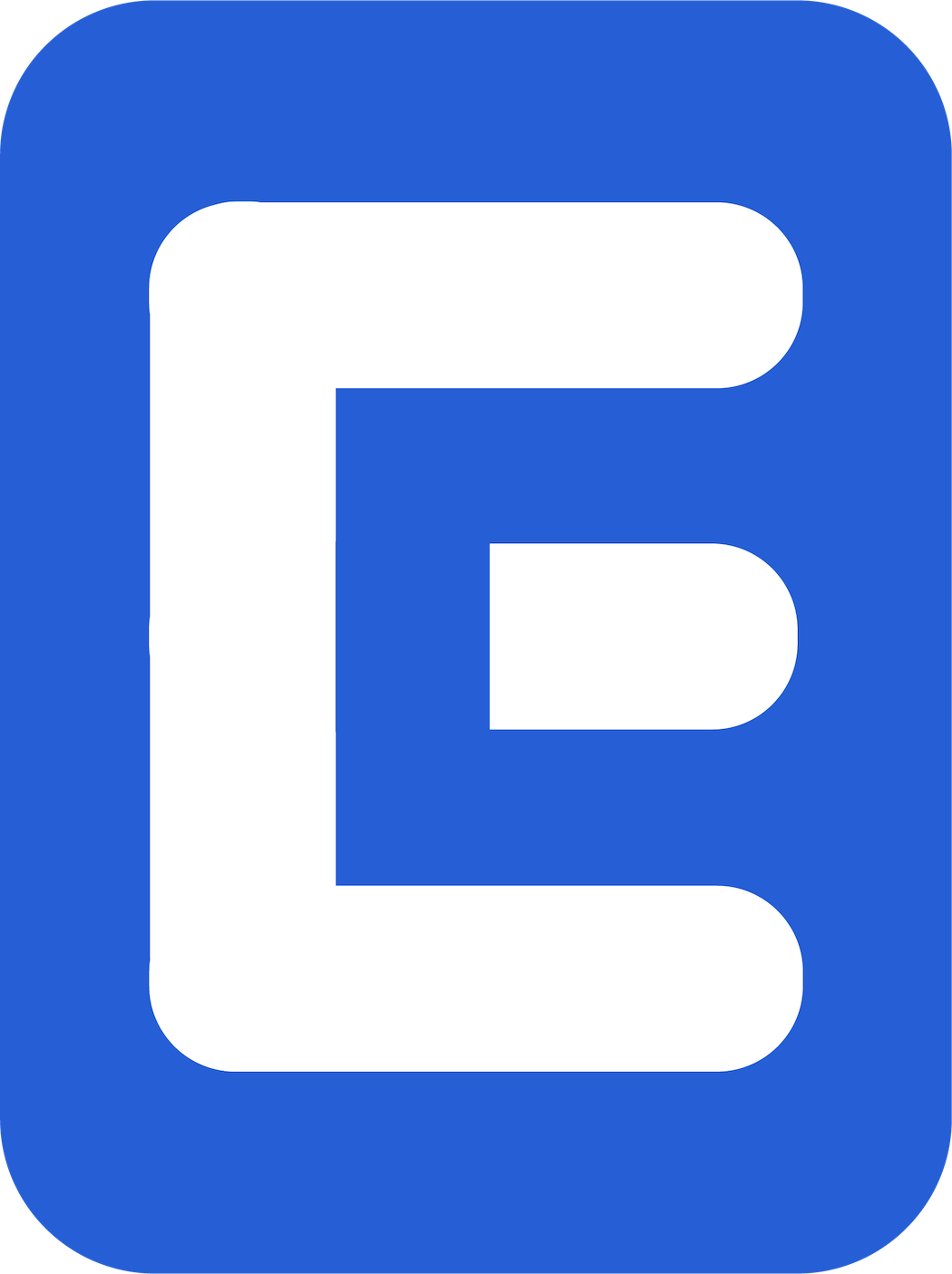}}

\makeatletter
\def\@maketitle{%
  \newpage
  \null
  \vskip 0.5em%
  \begin{center}%
  \let \footnote \thanks
    {\bfseries\Large \@title \par}%
    \vskip 1em%
    {\large
      \lineskip .5em%
      \begin{tabular}[t]{c}%
        \@author
      \end{tabular}\par}%
    \vskip 0.5em
  \end{center}%
  \par
  }
\makeatother

\title{\EndocIcon \textsc{ Modoc}: A Modular Interface for Flexible Interlinking of Text Retrieval and Text Generation Functions}

\author{$\textbf{Yingqiang Gao}^\diamond$, $\textbf{Jhony Prada}^\circ$, $\textbf{Nianlong Gu}^\vartriangle$, $\textbf{Jessica Lam}^\diamond$\\ $\textbf{Richard H.R. Hahnloser}^\diamond$ \\ 
$^\diamond\text{Institute of Neuroinformatics, University of Zurich and ETH Zurich, Switzerland}$ \\ 
\texttt{\{yingqiang.gao, lamjessica, rich\}@ini.ethz.ch} \\
$^\circ\text{ETH Zurich, Switzerland}$ \\
\texttt{jhonyp@ethz.ch} \\
$^\vartriangle\text{Linguistic Research Infrastructure, University of Zurich, Switzerland}$ \\ 
\texttt{nianlong.gu@uzh.ch} }

\begin{document}

\maketitle

\begin{abstract}

Large Language Models (LLMs) produce eloquent texts but often the content they generate needs to be verified. Traditional information retrieval systems can assist with this task, but most systems have not been designed with LLM-generated queries in mind. As such, there is a compelling need for integrated systems that provide both retrieval and generation functionality within a single user interface. 

We present \textsc{Modoc}, a modular user interface that leverages the capabilities of LLMs and provides assistance with detecting their confabulations, promoting integrity in scientific writing.
\textsc{Modoc} represents a significant step forward in scientific writing assistance. Its modular architecture supports flexible functions for retrieving information and for writing and generating text in a single, user-friendly interface.


\end{abstract}

\section{Introduction}

Scientific writing is a cognitively demanding task centered on formulating novel claims and discussing their relationship with published facts \cite{hayes2012modeling}. 
While modern information retrieval systems are highly effective in retrieving relevant documents from a given query \cite{fadaee2020new, hambarde2023information, zhu2023large}, users' requirements frequently extend past basic retrieval, e.g. when they wish to retrieve facts relevant to their manuscript without having to formulate a query or when adapting the retrieved content to their own narrative. 
Large Language Models (LLMs) can help with tasks such as narrative generation \cite{wadden2020fact, atanasova2020generating, smeros2021sciclops, gu2024controllable}, but they are prone to confabulations, which implies that their outputs must be scrutinized by a human and evaluated against the literature. This closed loop of desired retrieval and generation functionality emphasizes the necessity of integrating LLMs into scientific writing and retrieval workflows, to ease the workload and to allow users to concentrate on their science without being sidetracked by incongruent software interfaces.

\begin{figure}
    \includegraphics[width=\columnwidth]{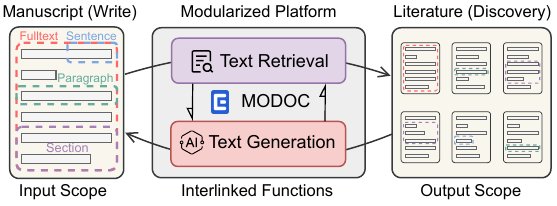}
    \caption{Interlinked Retrieval and Generation functions in our proposed platform \textsc{Modoc}. By setting the scope for input and output, the power of LLMs can be maximized for joint retrieval and generation. }
    \label{fig:teaser-new}
\end{figure}
Even though LLMs produce impressive scientific texts \cite{pan2021zero, wright2022generating, tan2023multi2claim}, the improper use of content generated by LLMs can lead to potential ethical concerns \cite{weidinger2021ethical, li2023ethics, zhuo2023exploring}. Inserting unverified LLM-generated content into manuscript drafts is dangerous because it can lead to the dissemination of inaccurate or misleading information, compromising the integrity and reliability of scientific research \cite{meyer2023chatgpt}. 
Therefore, scientific writing assistant systems should be designed to remove hurdles towards convenient verification of LLM-generated content.

In light of this state of affairs, we argue that there are two main issues that need to be addressed in modern scientific writing assistant systems: 1) interlinking retrieval and generation functions in a single user interface; 2) active guidance towards ethical usage of LLM-generated content through meticulously designed workflows. 

We introduce \textsc{Modoc} (see Figure~\ref{fig:teaser-new}), our solution for seamless integration of retrieval and generation of scientific content under a single modular user interface. \textsc{Modoc} is based on five modules, each of which can support up to a single interlinkable text retrieval or generation function. 
With \textsc{Modoc}'s modular components, users can effortlessly: 1) retrieve relevant scientific content on any level of granularity from millions of scientific documents in real time;
2) generate scientific texts with LLMs using freely definable contextualized prompts.

With \textsc{Modoc} modules, we aim to utilize the powers of LLMs and retrieval engines to promote factual and ethical scientific writing, while alleviating the associated cognitive burden. We believe that \textsc{Modoc} is the first practical attempt to enhance productivity in scientific writing rooted in flexible text retrieval and generation.
    
A live demo webpage of \textsc{Modoc} is at \url{https://endoc.ethz.ch}. We have recorded a video demonstration of \textsc{Modoc} which can be accessed at \url{https://youtu.be/--6URklQ27E}. The detailed guideline of \textsc{Modoc} can be found at \url{https://github.com/CharizardAcademy/modoc}.

\begin{figure*}
    \centering
\includegraphics[width=\textwidth]{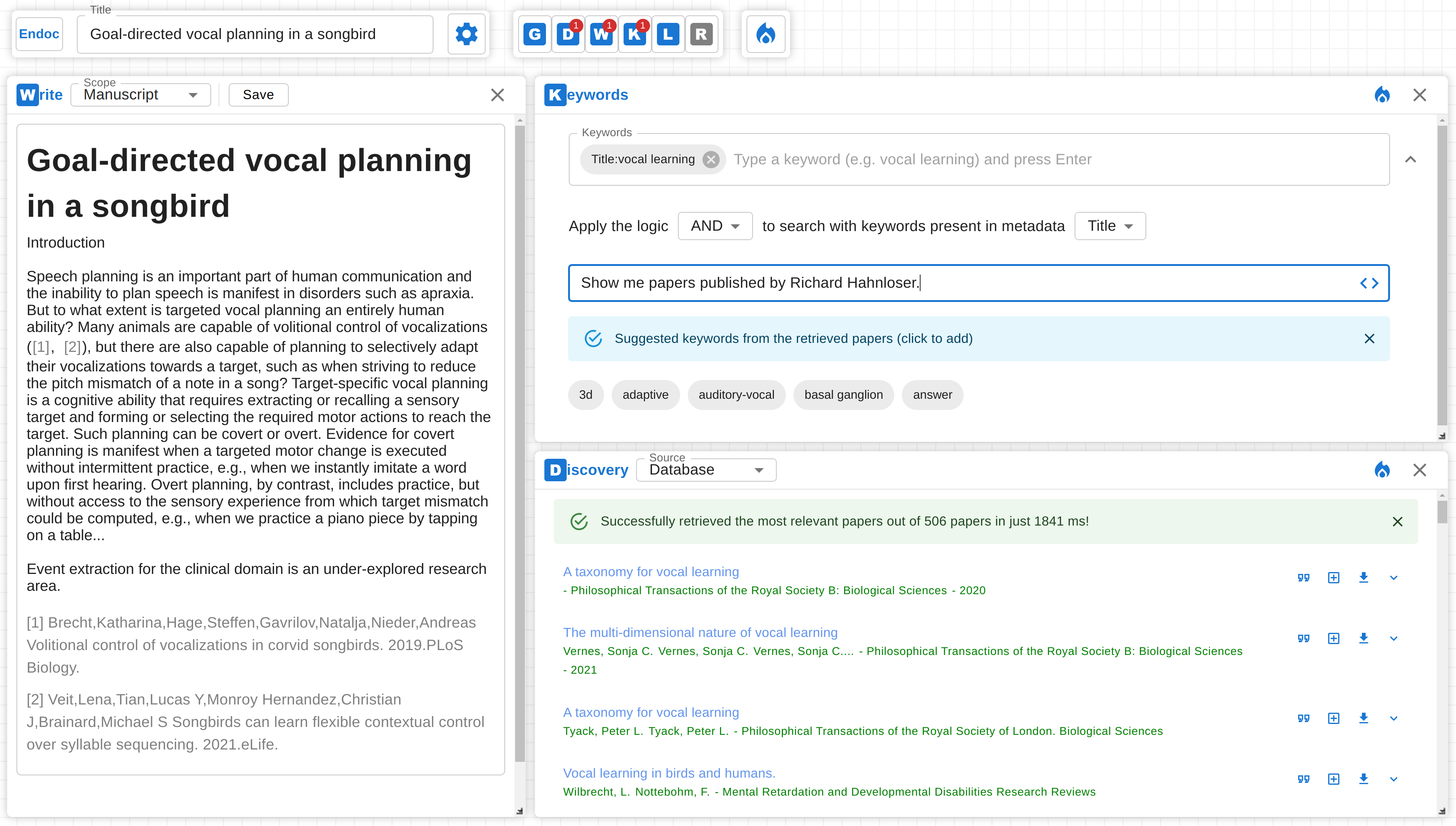}
    \caption{Overview of \textsc{Modoc}. This figure demonstrates the basic workflow Retrieve and Cite (detailed description in section~\ref{section:retrieve&cite}), where the user retrieves the most relevant research papers using keywords and the actual content of the manuscript. The modularity of \textsc{Modoc} ensures flexible configurations of many workflows not only for literature search, but also for verification of scientific claims and facts. The context within the Write module is taken from the work of \citet{zai2022goal}.}
    \label{fig:endoc-overview}
\end{figure*}

\section{Related Works}




Based on their features, current scientific writing assistant systems can be classified into three categories:

\begin{itemize}[left=0pt]
    \item \textbf{Retrieval systems that lack an integrated text editor}. Examples are \citet{aminer}, \citet{hypothesis}, \citet{paperpile}, \citet{iris-ai}, \citet{scispace}, \citet{zeta-alpha}, \citet{sciwheel}, \citet{scite}, \citet{deepsearch}, \citet{connected-papers}, \citet{scholarcy}, \citet{elicit}, etc. 
    Although these systems display sophisticated retrieval capabilities, they function merely as search engines and are unable to establish a direct connection with the user's manuscript. As a consequence, these systems leave an interlinking gap that 
    diverts authors' attention from the scientific content itself.    
    \item \textbf{Writing assistant systems that lack document retrieval capabilities}. Examples are \citet{grammarly}, \citet{paperpal}, \citet{instatext}, \citet{sciflow}, \citet{jargon}, etc.
    These solutions help with refining text but they provide no support for semantic queries, leaving extra work for linking the manuscript with pertinent information from external databases. 
    \item \textbf{Systems that do not provide guidance to users on the ethical use of LLM outputs}. Examples are \citet{quillbot}, \citet{authorea}, \citet{conch}, and \citet{jenniai}, etc. 
    Initially, these systems require users to manually restart the literature discovery process each time. Although they include a text editor and retrieval functions, their capability in producing scientific texts based on user inputs is not yet clear. Currently, users are unable to freely generate scientific texts from unstructured parts of their manuscript or from a paper they are currently reading.
\end{itemize}


\section{\textsc{Modoc}: Interlinking Retrieval and Generation Functions to Promote Specific Workflows}

Prior to \textsc{Modoc}, we developed two scientific writing assistant systems:
\begin{itemize}[left=0pt]
    \item \textsc{Endoc} \cite{gokcce2020embedding}, our first web-based text processing platform with literature discovery functionality. \textsc{Endoc}'s retrieval function ran on millions of fully indexed scientific papers and made use of both Boolean keyword filtering and embedding-based nearest neighbor search. 
    \item \textsc{SciLit} \cite{gu2023scilit} featured a rudimentary text processor that we interfaced with all the retrieval functions of \textsc{Endoc} and that incorporated advanced text generation pipelines such as citation sentence generation \cite{gu2024controllable} and document summarization \cite{gu2022memsum}. However, unlike \textsc{Modoc}, \textsc{SciLit} is a hardcoded demo interface and was not built from a vision to support retrieval and generation functions interlinkably in flexible ways.
\end{itemize}

Inheriting the rich text generation functionalities from \textsc{SciLit}, \textsc{Modoc} opts to provide ethical use of LLM-generated content by giving users the freedom of self-customizing the workflows. 


\subsection{Functions}

\textsc{Modoc} supports both \textit{retrieval} and \textit{generation} functions. We motivate the distinction of these function types by considering their trustworthiness. Generation functions are \textbf{abstractive} and creative, they can e.g. generate claims, but that also means they are at risk of confabulating. In contrast, retrieval functions are \textbf{extractive}, which means they never confabulate and so can assist with checking claims made by generation functions. By enforcing a clean separation of these function types, we promote constant awareness of where confabulations could enter the scientific writing process, allowing users to take mitigating actions.

The inputs and outputs of retrieval and generation functions are detailed in the following. 

\paragraph{Retrieval} Based on a query, retrieve a text at flexible levels of 
documents, sections, paragraphs, sentences, or keyphrases (keywords). The example retrieval functions currently supported are (\textit{literature}) \textit{discovery} (i.e., fetching the most pertinent documents), \textit{text alignment} (i.e., identifying the most pertinent text fragments within a specific document), and \textit{keyphrase (keywords) suggestion} (i.e., distilling relevant keyphrases from the 100 top documents for query refinement).

\begin{itemize}[left=0pt]
    \item \textbf{(Literature) discovery}: 
    Building upon our previous work \textsc{SciLit} \cite{gu2023scilit}, we harness the capability to  search through millions of documents in milliseconds while achieving high recall performance.
    \item \textbf{Text alignment}:  Inspired by Large-scale Hierarchical Alignment (LHA) introduced by \citet{nikolov2019large}, we frame the text alignment task as retrieving the most semantically relevant sentence within a document to the query text given by a user. In \textsc{Modoc}, the query text can be either user-selected or LLM-generated. Given a query text, we calculate the cosine similarity score between its Sent2vec \cite{pgj2017unsup} embedding and the embeddings of all other text pieces within the document, offering a wide range of text granularities encompassing sentences, paragraphs, and sections.
    \item \textbf{Keyphrases (keywords) suggestion}: 
    From each literature discovery, we identify five prepresentative keyphrases from the top 100 documents returned by the literature discovery tool. This is achieved by employing KeyBERT \cite{grootendorst2020keybert}, an embedding-based approach that identifies and extracts the most resonant keywords and keyphrases, as determined by the highest cosine similarity to the documents' vector representations.
\end{itemize}

\paragraph{Generation} Based on a prompt text along with contexts such as keywords and user-selected text, the generation functions generate an abstractive textual response. The currently supported functions are generation of a \textit{citation/conclusion sentence}. 

\begin{itemize}[left=0pt]
    \item \textbf{Citation sentence generation}: 
    We have incorporated our citation sentence generator SciCCG \cite{gu2024controllable} into \textsc{Modoc}. This integration produces citation sentences for inclusion in a manuscript.
    \item \textbf{Conclusion sentence generation}: We deployed \textsc{arg-align} \citep{gao2024evaluating}, our trained LLMs such as LlaMA \cite{touvron2023llama} and Galactica \cite{taylor2022galactica} produce conclusion sentences based on user-selected premises. 
\end{itemize}

To specify a function and its input sources, a user must wire it up with directional links among modules, as detailed in the following.

\subsection{Modules}
\textsc{Modoc}'s modules 
host the inputs and outputs of the configured functions. Once functions are wired up via links (a link means that one module provides input to another) and a desired workflow is established, users can change the textual inputs in one or several modules using the keyboard or mouse before triggering function calls. 

 \textsc{Modoc} comprises four modules for retrieval and one module for generation (see Table~\ref{tab:modules}).
 

\newcommand\cincludegraphics[2][]{\raisebox{-0.4\height}{\includegraphics[#1]{#2}}}

\begin{table}[!htb]
    \centering
    \resizebox{0.48\textwidth}{!}{
    \begin{tabular}{clc}
    \toprule
    Module & \multicolumn{1}{c}{Description} & Runs\\
    \midrule
    \multicolumn{1}{c}{\cincludegraphics[width=0.16\textwidth]{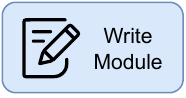}}  & \multirow{1}{*}{Text editor for writing a scientific manuscript.} & Ret \\
    \midrule
    \multicolumn{1}{c}
    {\cincludegraphics[width=0.16\textwidth]{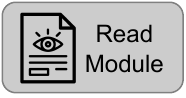}} & \pbox{7.8cm}{Module for reading scientific documents together with the highlights.} & Ret \\
    \midrule
    \multicolumn{1}{c}
    {\cincludegraphics[width=0.16\textwidth]{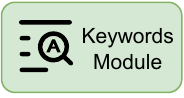}} & \pbox{7.8cm}{Module for specifying lexical filters 
 (keywords) for retrieval and generation.} & Ret \\
    \midrule
    \multicolumn{1}{c}
    {\cincludegraphics[width=0.16\textwidth]{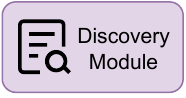}} & \pbox{7.8cm}{A literature discovery engine that retrieves the most relevant documents from the data servers.} & Ret \\
    \midrule
    \multicolumn{1}{c}
    {\cincludegraphics[width=0.16\textwidth]{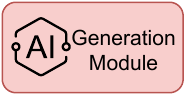}} & \pbox{7.8cm}{Module with integrated NLP services for generation.} & Gen \\
    \bottomrule
    \end{tabular}
    }
    \caption{Overview of the modules of \textsc{Modoc} and the functions they support (Ret=\textbf{Ret}rieval, Gen=\textbf{Gen}eration). See Appendix~\ref{sec:appendix-module} for more module details and Figure~\ref{fig:teaser} for module interactions.}
    \label{tab:modules}
\end{table}

\paragraph{Keywords} This module allows users to specify keyphrases (keywords) as inputs to NLP functions and it supports displaying keyphrases extracted by retrieval functions. Multiple keyphrases can be combined using Boolean logic to create a more complex filter, e.g. for literature discovery. 

\begin{figure}[!htb]
    \centering
    \includegraphics[width=\columnwidth]{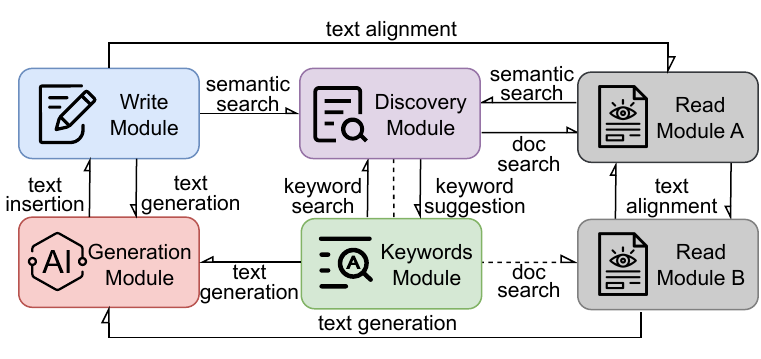}
    \caption{Interaction between the retrieval and generation modules. \textsc{Modoc} allows flexible configuration of these modules for certain workflows.}
    \label{fig:teaser}
\end{figure}

\paragraph{Discovery} This module displays the outcome of literature discovery, which is a retrieval function. For each document, its essential metadata (title, authors, venue, and publication date) is displayed. Four buttons enable users to: 1) Insert a citation into their manuscript; 2) Add the document to their personal library; 3) Download a reference of the document; and 4) Access the document's abstract.

\paragraph{Read} This module displays the text contained in a document, allowing users to select specific sections (scope). The currently supported retrieval functions are summarization (extract sentence highlights) and text alignment (extract sentences/paragraphs aligned with a given source text). Retrieved texts within the chosen scope can be displayed either in a list or as highlights in the document.

\paragraph{Write} 
This module comprises a text editor for drafting a manuscript. The list of references is displayed at the end of the document. 
The module can serve as text source for retrieving the most relevant text passage from the document currently being read.

\paragraph{Generation}
This module is destined for abstractive text generation, e.g., citation generation and conclusion generation. Citation generation is to generate a sentence within the context of the manuscript that refers to the chosen document  that matches a desired citation intent. Conclusion generation is to generate a sentence serving as inference of the selected premises.

As a design principle, for enforcing a given workflow, the output of a function can change the content of only the single module it is tied to, not several modules. To limit the complexity of the resulting workflows, functions cannot be linked in closed loops (to avoid endless looping of function calls). 

The benefits of \textsc{Modoc} are that after creating a workflow, it can be effortlessly re-executed multiple times as the underlying query arguments change.


\section{Workflows}
Using these modules, we exemplify workflows for citing a document and for adding LLM-generated text to a manuscript. In particular, we showcase variants of these workflows with increased ethical standards compared to the traditional norms of such activities.



\subsection{Retrieve and Cite}
\label{section:retrieve&cite}

\begin{figure}[!htb]
    \centering
    \includegraphics[width=\columnwidth]{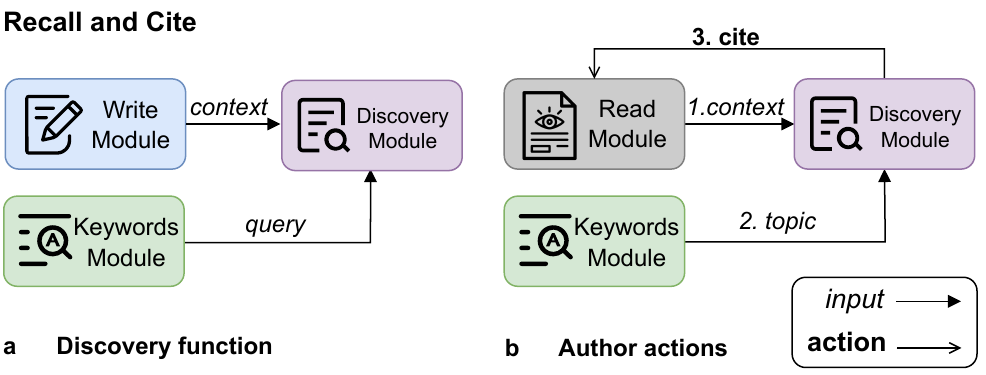}
    \caption{Recall and Cite workflow: (a) Configuration of the Discovery function. The Discovery module takes as input the query from the Keywords module and the context (i.e. the claim) from the manuscript (Write module); (b) Required author actions to perform Recall and Cite in chronological order.}
    \label{fig:recall-cite}
\end{figure}

Suppose that authors want to cite a document. Depending on the authors' intent and knowledge, we exemplify three variants of the Retrieve and Cite workflow: 

\paragraph{Recall and Cite (Figure~\ref{fig:recall-cite})} Authors cite a document without needing to read it (possibly because they have read the document before and they recall its content from memory). They select the citation context in the manuscript, execute literature discovery, and from the list of retrieved documents in the Discovery module cite the desired document. 

\begin{figure}[!htb]
    \centering
    \includegraphics[width=\columnwidth]{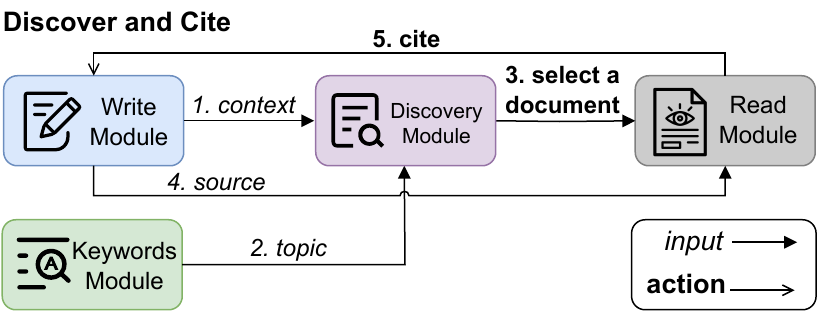}
    \caption{Discover and Cite workflow. The required author actions to perform this workflow are shown in chronological order.}
    \label{fig:discover-cite}
\end{figure}

\paragraph{Discover and Cite (Figure~\ref{fig:discover-cite})} Authors first perform a search for a suitable document, then read it, and finally cite it. They read a candidate document in the Read module, and when they find a suitable document, cite it in the manuscript from within the Read module.

\begin{figure}[!htb]
    \centering
    \includegraphics[width=\columnwidth]{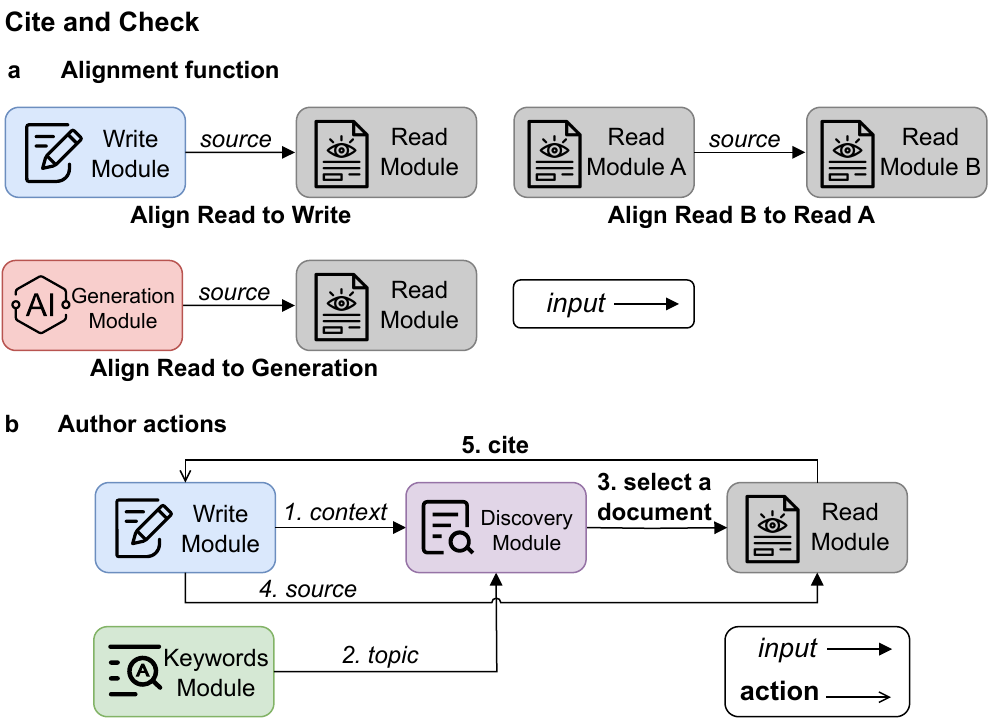}
    \caption{Cite and Check workflow. (a) Authors perform text alignment based on either a text in the manuscript (Write module), a text in a second Read module, or a text generated in a Generation module. The alignment function retrieves the most relevant text in the document; (b) Required author actions to perform Check and Cite in chronological order.}
    \label{fig:check-cite}
\end{figure}

\paragraph{Cite and Check (Figure~\ref{fig:check-cite})} To enhance the ethical standard of Recall/Discover and Cite, a simple check is advisable. To check the citation text for consistency with the cited document (e.g. in case the document is very long and may contain conflicting information the user overlooked), authors select the manually generated citation sentence, and use it as a query for text alignment against the document. The most similar statements are then highlighted. 
Additionally, they may want to check whether the citation agrees with the wider literature, which they can do by using the citation sentence as a source for document discovery, etc.

To configure these three Retrieve and Cite workflows, authors performs the following steps: 
\begin{itemize}[left=0pt]
    \item Authors set up the Discovery module and its retrieval function to expect an input from the Write module;
    \item Optionally, they configure the Discovery module to also expect lexical filters from the Keywords module;
    \item Optionally, they configure the Read module to perform text alignment against input from the Write module.
\end{itemize}

\subsection{Generate and Check}

Assume authors need to formulate a statement (e.g. a claim or hypothesis) in the manuscript and wish to verify its novelty/accuracy by searching published documents with related information potentially worth citing. They seek help from an LLM to formulate the statement. 

\begin{figure}[!htb]
    \centering
    \includegraphics[width=\columnwidth]{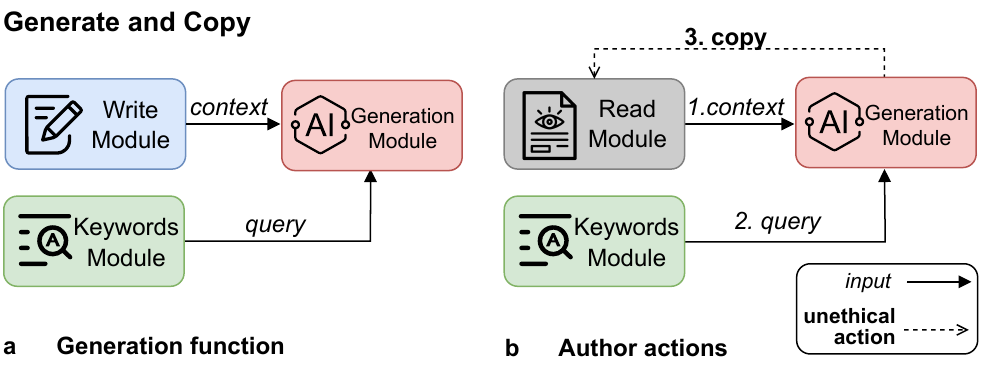}
    \caption{Generate and Copy. (a) Configuration of the Generation function that takes input from the manuscript (Write Module) and optionally from the Keywords Module; (b) Author actions required to perform claim generation, in chronological order. Directly copying the generated claim into the manuscript without checking (dashed line in Figure) is a potentially unethical use of LLM output.}
    \label{fig:generate-copy}
\end{figure}

Based on the authors' intention, we present the following two variants of this workflow:
\paragraph{Generate and Copy (Figure~\ref{fig:generate-copy})} Based on the selected context, the LLM generates a claim that authord directly adopt in the manuscript. 
This workflow is appropriate when authors are sufficiently knowledgeable about the generated claim without further checking. However, the non-critical use of the text generated by LLMs poses a significant risk of producing unreliable scientific content.

\begin{figure}[!htb]
    \centering
    \includegraphics[width=\columnwidth]{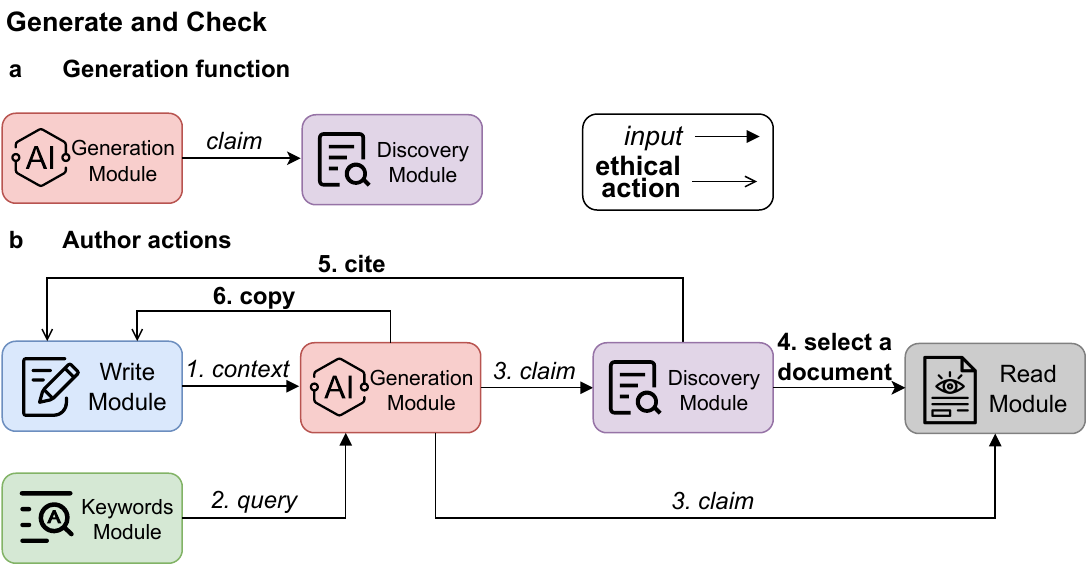}
    \caption{Generate and Check. (a) In addition to the Generate function, authors configure also a Discovery module, expecting input from the Generation module. In this ethical workflow, they first check the claim via literature discovery and text alignment; (b) Required author actions to perform Generation and Check in chronological order.}
    \label{fig:generate-check}
\end{figure}

\paragraph{Generate and Check (Figure~\ref{fig:generate-check})} The more ethical variant of this workflow is to check the LLM-generated claim against the literature before adopting it in the manuscript. This workflow requires the additional configuration of a retrieval function. 

To configure the two Generate and Check workflows, authors perform the following steps: 
\begin{itemize}[left=0pt]
    \item Authors configure the Generation module and its generation function to expect input from the Write or Read module;
    \item Optionally, they configures the Generation module to expect constraining keywords from the Keywords module;
    \item To check, they configure the Discovery module to expect input from the Generation module.
\end{itemize} 

The workflows inherent to \textsc{Modoc} are intricately aligned with the evolving dynamics of a user's interests. \textsc{Modoc} is designed to monitor and adapt to changes in text selection and the alteration of keywords by users. This ensures a smooth integration of NLP services for execution at any given moment. This approach underscores \textsc{Modoc}'s commitment to providing responsive and adaptive NLP solutions.

\section{Conclusions}

This demo introduces \textsc{Modoc}, a modularized prototype for integrated retrieval, generation, and verification during scientific writing. \textsc{Modoc} leverages highly flexible workflow configuration to advance content retrieval across various levels of granularity. It aids users in engaging with LLMs, allowing them to harness the capabilities of state-of-the-art language models while mitigating the risk of incorporating unverified content into their manuscripts.
The benefit of \textsc{Modoc} is it allows users to define flexible workflows for repetitive writing tasks helping users shift their attention from distractions to the contents of their science.

\section{Limitations}

As an early prototype, \textsc{Modoc} comes with non-polished designs. In future work, we plan to undertake comprehensive user studies on \textsc{Modoc} to refine our system design. 
This includes exploring automatic workflow recommendations based on natural language instructions and probabilistic ethics checks for manuscript content.
We are also committed to enhancing the core functionalities of 
\textsc{Modoc}, focusing on advanced retrieval and generation functions for ethical workflows.

\section{Ethics Statement} 

The text generation capabilities of \textsc{Modoc}, specifically those pertaining to the generation of citation sentences and conclusion sentences, possess the potential to produce content that lacks fidelity and accuracy. Such deviations from faithfulness and factuality may cause unethical use of content generated by LLMs. We, therefore, emphasize the necessity of adhering to recommended ethical workflows in order to safeguard the integrity and ethical utilization of content produced by LLMs.

\section*{Acknowledgements}

We thank our former colleagues Dr. Onur G\"okçe, Dr. Nikola I. Nikolov, and Dr. Andrei Plamada for their contributions to \textsc{Modoc}.

We acknowledge the support from Swiss National Science Foundation NCCR Evolving Language, Agreement No.51NF40\_180888, as well as Swiss Open Research Data Grants, number ORD2000103. 
\bibliography{custom}

\begin{thebibliography}{49}
\expandafter\ifx\csname natexlab\endcsname\relax\def\natexlab#1{#1}\fi

\bibitem[{Achiam et~al.(2023)Achiam, Adler, Agarwal, Ahmad, Akkaya, Aleman, Almeida, Altenschmidt, Altman, Anadkat et~al.}]{achiam2023gpt}
Josh Achiam, Steven Adler, Sandhini Agarwal, Lama Ahmad, Ilge Akkaya, Florencia~Leoni Aleman, Diogo Almeida, Janko Altenschmidt, Sam Altman, Shyamal Anadkat, et~al. 2023.
\newblock \href {https://arxiv.org/abs/2303.08774} {Gpt-4 technical report}.
\newblock \emph{arXiv preprint arXiv:2303.08774}.

\bibitem[{{Aminer}()}]{aminer}
{Aminer}.
\newblock \href {https://www.aminer.cn/} {{Aminer} - academic search engine}.

\bibitem[{Atanasova et~al.(2020)Atanasova, Simonsen, Lioma, and Augenstein}]{atanasova2020generating}
Pepa Atanasova, Jakob~Grue Simonsen, Christina Lioma, and Isabelle Augenstein. 2020.
\newblock Generating fact checking explanations.
\newblock In \emph{Proceedings of the 58th Annual Meeting of the Association for Computational Linguistics}, pages 7352--7364.

\bibitem[{{Authorea}()}]{authorea}
{Authorea}.
\newblock \href {https://www.authorea.com/} {{Authorea} - collaborative writing and publishing platform}.

\bibitem[{{Conch}()}]{conch}
{Conch}.
\newblock \href {https://www.getconch.ai/} {{Conch} - ai-powered writing and editing}.

\bibitem[{{Connected Papers}()}]{connected-papers}
{Connected Papers}.
\newblock \href {https://www.connectedpapers.com/} {{Connected Papers} - explore papers and discover connections}.

\bibitem[{{Elicit}()}]{elicit}
{Elicit}.
\newblock \href {https://elicit.com/} {{Elicit} - online survey and market research tool}.

\bibitem[{Fadaee et~al.(2020)Fadaee, Gureenkova, Barrera, Schnober, Weerkamp, and Zavrel}]{fadaee2020new}
Marzieh Fadaee, Olga Gureenkova, Fernando~Rejon Barrera, Carsten Schnober, Wouter Weerkamp, and Jakub Zavrel. 2020.
\newblock A new neural search and insights platform for navigating and organizing ai research.
\newblock In \emph{Proceedings of the First Workshop on Scholarly Document Processing}, pages 207--213.

\bibitem[{Gao et~al.(2024)Gao, Gu, Lam, Henderson, and Hahnloser}]{gao2024evaluating}
Yingqiang Gao, Nianlong Gu, Jessica Lam, James Henderson, and Richard Hahnloser. 2024.
\newblock Evaluating unsupervised argument aligners via generation of conclusions of structured scientific abstracts.
\newblock In \emph{Proceedings of the 18th Conference of the European Chapter of the Association for Computational Linguistics (Volume 2: Short Papers)}, pages 151--160.

\bibitem[{G{\"o}k{\c{c}}e et~al.(2020)G{\"o}k{\c{c}}e, Prada, Nikolov, Gu, and Hahnloser}]{gokcce2020embedding}
Onur G{\"o}k{\c{c}}e, Jonathan Prada, Nikola~I Nikolov, Nianlong Gu, and Richard~HR Hahnloser. 2020.
\newblock Embedding-based scientific literature discovery in a text editor application.
\newblock In \emph{Proceedings of the 58th Annual Meeting of the Association for Computational Linguistics: System Demonstrations}, pages 320--326.

\bibitem[{{Grammarly}()}]{grammarly}
{Grammarly}.
\newblock \href {https://www.grammarly.com/} {{Grammarly} - writing assistant}.

\bibitem[{Grootendorst()}]{grootendorst2020keybert}
Maarten Grootendorst.
\newblock \href {https://doi.org/10.5281/zenodo.4461265} {Keybert: Minimal keyword extraction with bert.}

\bibitem[{Gu et~al.(2022)Gu, Ash, and Hahnloser}]{gu2022memsum}
Nianlong Gu, Elliott Ash, and Richard Hahnloser. 2022.
\newblock Memsum: Extractive summarization of long documents using multi-step episodic markov decision processes.
\newblock In \emph{Proceedings of the 60th Annual Meeting of the Association for Computational Linguistics (Volume 1: Long Papers)}, pages 6507--6522.

\bibitem[{Gu and Hahnloser(2024)}]{gu2024controllable}
Nianlong Gu and Richard Hahnloser. 2024.
\newblock Controllable citation sentence generation with language models.
\newblock In \emph{Proceedings of the Fourth Workshop on Scholarly Document Processing (SDP 2024)}, pages 22--37.

\bibitem[{Gu and Hahnloser(2022)}]{gu2022controllable}
Nianlong Gu and Richard~HR Hahnloser. 2022.
\newblock Controllable citation text generation.
\newblock \emph{arXiv preprint arXiv:2211.07066}.

\bibitem[{Gu and Hahnloser(2023)}]{gu2023scilit}
Nianlong Gu and Richard~H.R. Hahnloser. 2023.
\newblock \href {https://doi.org/10.18653/v1/2023.acl-demo.22} {{S}ci{L}it: A platform for joint scientific literature discovery, summarization and citation generation}.
\newblock In \emph{Proceedings of the 61st Annual Meeting of the Association for Computational Linguistics (Volume 3: System Demonstrations)}, pages 235--246, Toronto, Canada. Association for Computational Linguistics.

\bibitem[{Hambarde and Proenca(2023)}]{hambarde2023information}
Kailash~A Hambarde and Hugo Proenca. 2023.
\newblock Information retrieval: Recent advances and beyond.
\newblock \emph{arXiv preprint arXiv:2301.08801}.

\bibitem[{Hayes(2012)}]{hayes2012modeling}
John~R Hayes. 2012.
\newblock Modeling and remodeling writing.
\newblock \emph{Written communication}, 29(3):369--388.

\bibitem[{{hypothes.is}()}]{hypothesis}
{hypothes.is}.
\newblock \href {https://web.hypothes.is/} {{hypothes.is} - annotate the web, with anyone, anywhere}.

\bibitem[{{IBM Deep Search}()}]{deepsearch}
{IBM Deep Search}.
\newblock \href {https://github.com/DS4SD/deepsearch-toolkit} {{Deep Search Toolkit}}.

\bibitem[{{InstaText}()}]{instatext}
{InstaText}.
\newblock \href {https://instatext.io/academic-writing/} {{InstaText} - academic writing assistant}.

\bibitem[{{Iris.ai}()}]{iris-ai}
{Iris.ai}.
\newblock \href {https://iris.ai/} {{Iris.ai} - the ai science assistant}.

\bibitem[{{Jargon}()}]{jargon}
{Jargon}.
\newblock \href {https://www.explainjargon.com/} {{Jargon} - simplify complex language and terms}.

\bibitem[{{Jenni.ai}()}]{jenniai}
{Jenni.ai}.
\newblock \href {https://jenni.ai} {Jenni ai: Your ai research assistant}.
\newblock [Online; accessed 15-January-2024].

\bibitem[{Jiang et~al.(2023)Jiang, Sablayrolles, Mensch, Bamford, Chaplot, Casas, Bressand, Lengyel, Lample, Saulnier et~al.}]{jiang2023mistral}
Albert~Q Jiang, Alexandre Sablayrolles, Arthur Mensch, Chris Bamford, Devendra~Singh Chaplot, Diego de~las Casas, Florian Bressand, Gianna Lengyel, Guillaume Lample, Lucile Saulnier, et~al. 2023.
\newblock Mistral 7b.
\newblock \emph{arXiv preprint arXiv:2310.06825}.

\bibitem[{Li et~al.(2023)Li, Moon, Purkayastha, Celi, Trivedi, and Gichoya}]{li2023ethics}
Hanzhou Li, John~T Moon, Saptarshi Purkayastha, Leo~Anthony Celi, Hari Trivedi, and Judy~W Gichoya. 2023.
\newblock Ethics of large language models in medicine and medical research.
\newblock \emph{The Lancet Digital Health}, 5(6):e333--e335.

\bibitem[{Meyer et~al.(2023)Meyer, Urbanowicz, Martin, O’Connor, Li, Peng, Bright, Tatonetti, Won, Gonzalez-Hernandez et~al.}]{meyer2023chatgpt}
Jesse~G Meyer, Ryan~J Urbanowicz, Patrick~CN Martin, Karen O’Connor, Ruowang Li, Pei-Chen Peng, Tiffani~J Bright, Nicholas Tatonetti, Kyoung~Jae Won, Graciela Gonzalez-Hernandez, et~al. 2023.
\newblock Chatgpt and large language models in academia: opportunities and challenges.
\newblock \emph{BioData Mining}, 16(1):20.

\bibitem[{Nikolov and Hahnloser(2019)}]{nikolov2019large}
Nikola~I Nikolov and Richard Hahnloser. 2019.
\newblock Large-scale hierarchical alignment for data-driven text rewriting.
\newblock In \emph{Proceedings of the International Conference on Recent Advances in Natural Language Processing (RANLP 2019)}, pages 844--853.

\bibitem[{Pagliardini et~al.(2018)Pagliardini, Gupta, and Jaggi}]{pgj2017unsup}
Matteo Pagliardini, Prakhar Gupta, and Martin Jaggi. 2018.
\newblock {Unsupervised Learning of Sentence Embeddings using Compositional n-Gram Features}.
\newblock In \emph{NAACL 2018 - Conference of the North American Chapter of the Association for Computational Linguistics}.

\bibitem[{Pan et~al.(2021)Pan, Chen, Xiong, Kan, and Wang}]{pan2021zero}
Liangming Pan, Wenhu Chen, Wenhan Xiong, Min-Yen Kan, and William~Yang Wang. 2021.
\newblock Zero-shot fact verification by claim generation.
\newblock In \emph{Proceedings of the 59th Annual Meeting of the Association for Computational Linguistics and the 11th International Joint Conference on Natural Language Processing (Volume 2: Short Papers)}, pages 476--483.

\bibitem[{{Paperpal}()}]{paperpal}
{Paperpal}.
\newblock \href {https://paperpal.com/} {{Paperpal} - ai-powered writing assistant}.

\bibitem[{{Paperpile}()}]{paperpile}
{Paperpile}.
\newblock \href {https://paperpile.com/} {{Paperpile} - reference management and citation software}.

\bibitem[{{QuillBot}()}]{quillbot}
{QuillBot}.
\newblock \href {https://quillbot.com/} {{QuillBot} - ai-powered writing assistant}.

\bibitem[{{Scholarcy}()}]{scholarcy}
{Scholarcy}.
\newblock \href {https://www.scholarcy.com/} {{Scholarcy} - ai-powered research tools}.

\bibitem[{{SciFlow}()}]{sciflow}
{SciFlow}.
\newblock \href {https://www.sciflow.net/en/} {{SciFlow} - collaborative research and writing platform}.

\bibitem[{{SciSpace}()}]{scispace}
{SciSpace}.
\newblock \href {https://typeset.io/} {{SciSpace} - research collaboration and manuscript writing platform}.

\bibitem[{{scite}()}]{scite}
{scite}.
\newblock \href {https://scite.ai/} {{scite} - discover and evaluate scientific articles}.

\bibitem[{{Sciwheel}()}]{sciwheel}
{Sciwheel}.
\newblock \href {https://sciwheel.com/?lg} {{Sciwheel} - reference management and collaboration for researchers}.

\bibitem[{Smeros et~al.(2021)Smeros, Castillo, and Aberer}]{smeros2021sciclops}
Panayiotis Smeros, Carlos Castillo, and Karl Aberer. 2021.
\newblock Sciclops: Detecting and contextualizing scientific claims for assisting manual fact-checking.
\newblock In \emph{Proceedings of the 30th ACM international conference on information \& knowledge management}, pages 1692--1702.

\bibitem[{Tan et~al.(2023)Tan, Nguyen, Bensemann, Peng, Bao, Chen, Gahegan, and Witbrock}]{tan2023multi2claim}
Neset Tan, Trung Nguyen, Josh Bensemann, Alex Peng, Qiming Bao, Yang Chen, Mark Gahegan, and Michael~J Witbrock. 2023.
\newblock Multi2claim: Generating scientific claims from multi-choice questions for scientific fact-checking.
\newblock In \emph{Proceedings of the 17th Conference of the European Chapter of the Association for Computational Linguistics}, pages 2644--2656.

\bibitem[{Taylor et~al.(2022)Taylor, Kardas, Cucurull, Scialom, Hartshorn, Saravia, Poulton, Kerkez, and Stojnic}]{taylor2022galactica}
Ross Taylor, Marcin Kardas, Guillem Cucurull, Thomas Scialom, Anthony Hartshorn, Elvis Saravia, Andrew Poulton, Viktor Kerkez, and Robert Stojnic. 2022.
\newblock Galactica: A large language model for science.
\newblock \emph{arXiv preprint arXiv:2211.09085}.

\bibitem[{Touvron et~al.(2023)Touvron, Lavril, Izacard, Martinet, Lachaux, Lacroix, Rozi{\`e}re, Goyal, Hambro, Azhar et~al.}]{touvron2023llama}
Hugo Touvron, Thibaut Lavril, Gautier Izacard, Xavier Martinet, Marie-Anne Lachaux, Timoth{\'e}e Lacroix, Baptiste Rozi{\`e}re, Naman Goyal, Eric Hambro, Faisal Azhar, et~al. 2023.
\newblock Llama: Open and efficient foundation language models.
\newblock \emph{arXiv preprint arXiv:2302.13971}.

\bibitem[{Wadden et~al.(2020)Wadden, Lin, Lo, Wang, van Zuylen, Cohan, and Hajishirzi}]{wadden2020fact}
David Wadden, Shanchuan Lin, Kyle Lo, Lucy~Lu Wang, Madeleine van Zuylen, Arman Cohan, and Hannaneh Hajishirzi. 2020.
\newblock Fact or fiction: Verifying scientific claims.
\newblock In \emph{Proceedings of the 2020 Conference on Empirical Methods in Natural Language Processing (EMNLP)}, pages 7534--7550.

\bibitem[{Weidinger et~al.(2021)Weidinger, Mellor, Rauh, Griffin, Uesato, Huang, Cheng, Glaese, Balle, Kasirzadeh et~al.}]{weidinger2021ethical}
Laura Weidinger, John Mellor, Maribeth Rauh, Conor Griffin, Jonathan Uesato, Po-Sen Huang, Myra Cheng, Mia Glaese, Borja Balle, Atoosa Kasirzadeh, et~al. 2021.
\newblock Ethical and social risks of harm from language models.
\newblock \emph{arXiv preprint arXiv:2112.04359}.

\bibitem[{Wright et~al.(2022)Wright, Wadden, Lo, Kuehl, Cohan, Augenstein, and Wang}]{wright2022generating}
Dustin Wright, David Wadden, Kyle Lo, Bailey Kuehl, Arman Cohan, Isabelle Augenstein, and Lucy~Lu Wang. 2022.
\newblock Generating scientific claims for zero-shot scientific fact checking.
\newblock In \emph{Proceedings of the 60th Annual Meeting of the Association for Computational Linguistics (Volume 1: Long Papers)}, pages 2448--2460.

\bibitem[{Zai et~al.(2022)Zai, Stepien, Cav{\'e}-Lopez, Giret, and Hahnloser}]{zai2022goal}
Anja~T Zai, Anna~E Stepien, Sophie Cav{\'e}-Lopez, Nicolas Giret, and Richard~HR Hahnloser. 2022.
\newblock Goal-directed vocal planning in a songbird.
\newblock \emph{bioRxiv}, pages 2022--09.

\bibitem[{{Zeta Alpha}()}]{zeta-alpha}
{Zeta Alpha}.
\newblock \href {https://www.zeta-alpha.com/} {{Zeta Alpha} - ai-driven content generation}.

\bibitem[{Zhu et~al.(2023)Zhu, Yuan, Wang, Liu, Liu, Deng, Dou, and Wen}]{zhu2023large}
Yutao Zhu, Huaying Yuan, Shuting Wang, Jiongnan Liu, Wenhan Liu, Chenlong Deng, Zhicheng Dou, and Ji-Rong Wen. 2023.
\newblock Large language models for information retrieval: A survey.
\newblock \emph{arXiv preprint arXiv:2308.07107}.

\bibitem[{Zhuo et~al.(2023)Zhuo, Huang, Chen, and Xing}]{zhuo2023exploring}
Terry~Yue Zhuo, Yujin Huang, Chunyang Chen, and Zhenchang Xing. 2023.
\newblock Exploring ai ethics of chatgpt: A diagnostic analysis.
\newblock \emph{arXiv preprint arXiv:2301.12867}.

\end{thebibliography}
\bibliographystyle{acl_natbib}

\appendix








\newpage
\onecolumn

\newcommand{\fire}[1]{
  \includegraphics[width=0.75em]{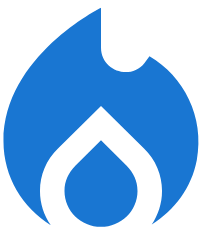}
}

\section{Detailed Module Description}

\label{sec:appendix-module}

\begin{figure}[!htb]
    \centering
    \includegraphics[width=\columnwidth]{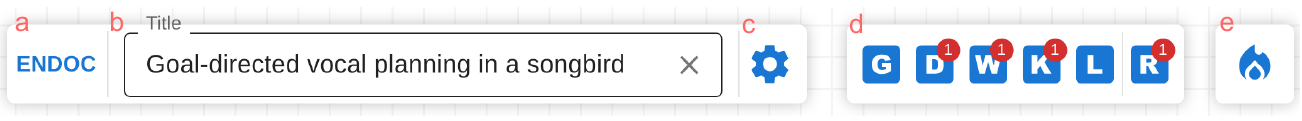}
    \caption{Overview of the \textsc{Modoc} Toolbar. (a) A ``Back'' button for returning to the manuscript selection page; (b) The title of the current manuscript, editable at any moment; (c) A button to revert to the classic interface and access account settings; (d) Dedicated buttons for each module, which open the corresponding module in a new window upon clicking; (e) A global ``Fire'' button\fire tthat activates all NLP service APIs, utilizing context from all modules.}
    \label{fig:toolbar-module}
\end{figure}

\paragraph{Toolbar (Figure~\ref{fig:toolbar-module})} 
The toolbar located at the top of the interface serves as a central hub for accessing both the modules and essential functions. It allows users to toggle between the classic (\textsc{Endoc}) and modular (\textsc{Modoc}) interfaces and to easily navigate back to the manuscript page, among other basic operations.

\begin{figure*}[!htb]
    \centering
    \includegraphics[width=\textwidth]{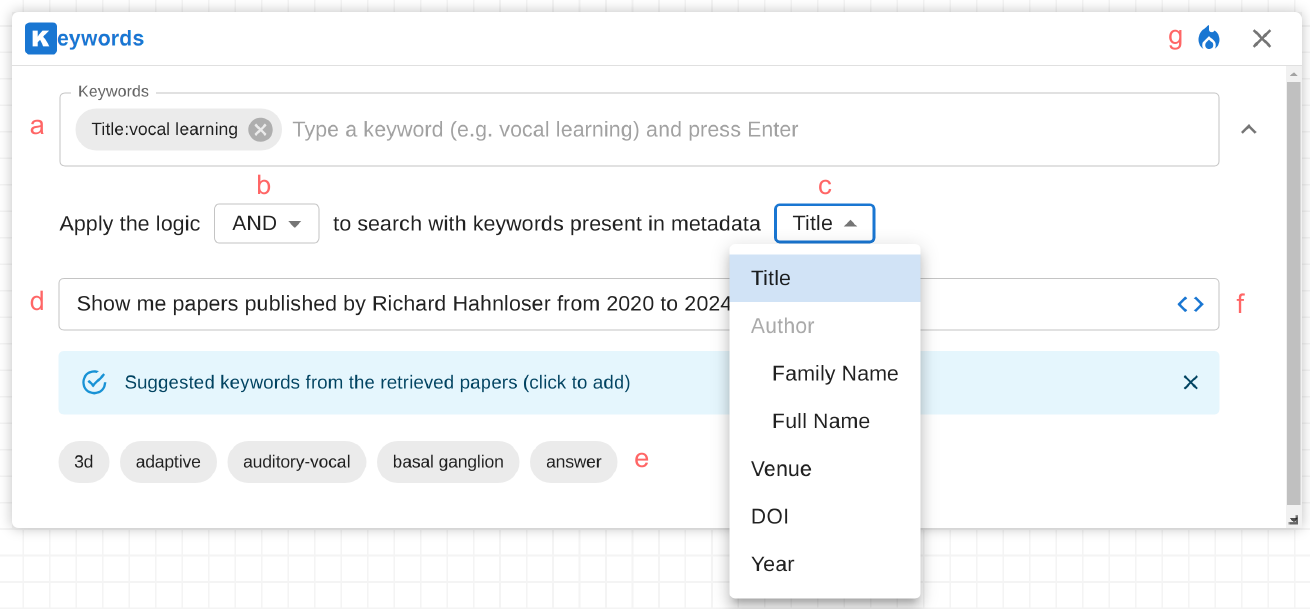}
    \caption{Overview of the Keywords module: (a) Literature discovery search bar: enter keywords as separate tabs for targeted searches; (b) Logic selection: choose the search logic for the current search item (AND logic is applied to all search items by default, whereas NOT logic has to be specified to a certain search item); (c) Prefix selection: choose prefixes to ensure search results include specific metadata. For example, \texttt{Title:vocal learning} filters documents to those with ``vocal learning'' in the title; (d) Search with natural language query: users can input a search query with their own words; (e) Suggested keywords: refine search results using keywords recommended based on the initial retrieval of 100 documents; (f) ``Parse'' button: click to parse the natural text query into a list of search items. We deploy a Mistral 7B \cite{jiang2023mistral} model to parse the natural text query (e.g., [\texttt{Author.FullName:Richard Hahnloser}, \texttt{Year:2020..2024}]); (g) Local ``Fire'' button\fire:: click to trigger the literature discovery.}
    \label{fig:keywords-module}
\end{figure*}

\paragraph{Keywords (Figure~\ref{fig:keywords-module})} A module that requires input in the form of keywords (which can consist of single or multiple text spans) or a semantic search query to initiate the process of literature discovery. 
The user initiates the process by entering keywords into the search bar. Upon pressing ENTER on the keyboard, each keyword is displayed in a separate tab, accompanied by a prefix chosen from a dropdown menu located below the input bar. This prefix denotes the specific metadata information required to be encompassed in the search results.
To omit specific keyword tabs from the search results, the user can activate the NOT checkbox. Additionally, the user can delineate the desired range of publication years by adjusting the start and end points on the horizontal scroll bar.

Subsequently, the user initiates the literature discovery process by clicking on the\fire icon situated in the upper right corner. After the system retrieves the initial set of 100 documents, it proposes an additional five keywords to refine the accuracy of the literature search results. The author can effortlessly incorporate these suggested keywords by selecting the corresponding tabs located below the search bar, which are then added as additional keyword tabs. 
Upon completion of the literature discovery process, the Discovery module will automatically open.

\begin{figure*}[!htb]
    \centering
    \includegraphics[width=\textwidth]{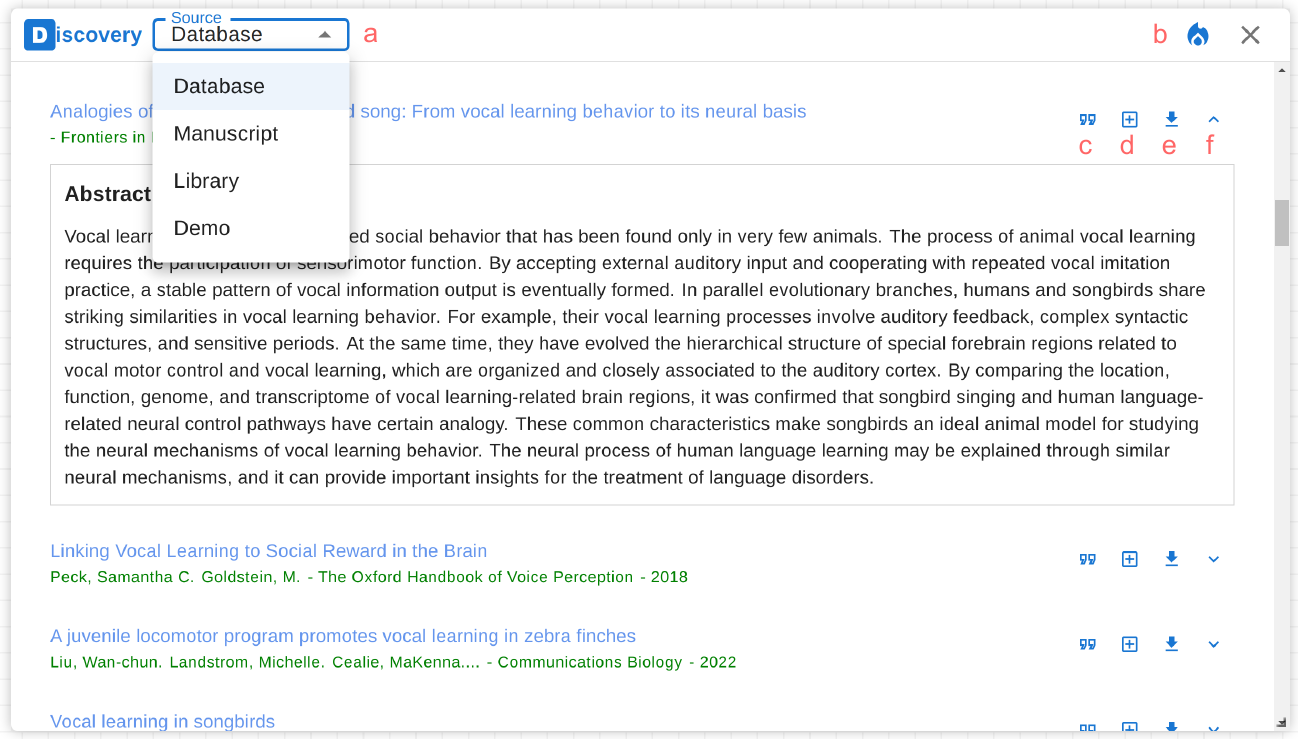}
    \caption{Overview of the Discovery module. (a) Retrieval source selection: choose \texttt{Database} (default) to display documents from the entire database matching the keywords. Selecting \texttt{Manuscript} initiates a semantic search using selected content from the Write module; (b) Local ``Fire'' button\fire:: click to view the retrieved documents; (c) Manuscript citation: add citations directly into user's draft in the Write module by placing a citation marker at the cursor position; (d) Library addition: save the document to the user’s personal library (Library module); (e) Reference formation: create and download a citation entry for the document in MLA or BibTex format; (f) Document card expansion/folding: toggle to show or hide the document's abstract.}
    \label{fig:discovery-module}
\end{figure*}

\paragraph{Discovery (Figure~\ref{fig:discovery-module})} This module presents literature discovery results based on keywords given in the Keywords Module. It displays each document as a document card, showcasing metadata such as \texttt{Title, Author, Venue}, and \texttt{Publication Date}. In addition, users can conduct a semantic search using selected content from user's draft (Write module) to find the most relevant documents. To utilize this feature, select \texttt{Manuscript} from the Source menu.

\begin{figure}
    \centering
    \includegraphics[width=\columnwidth]{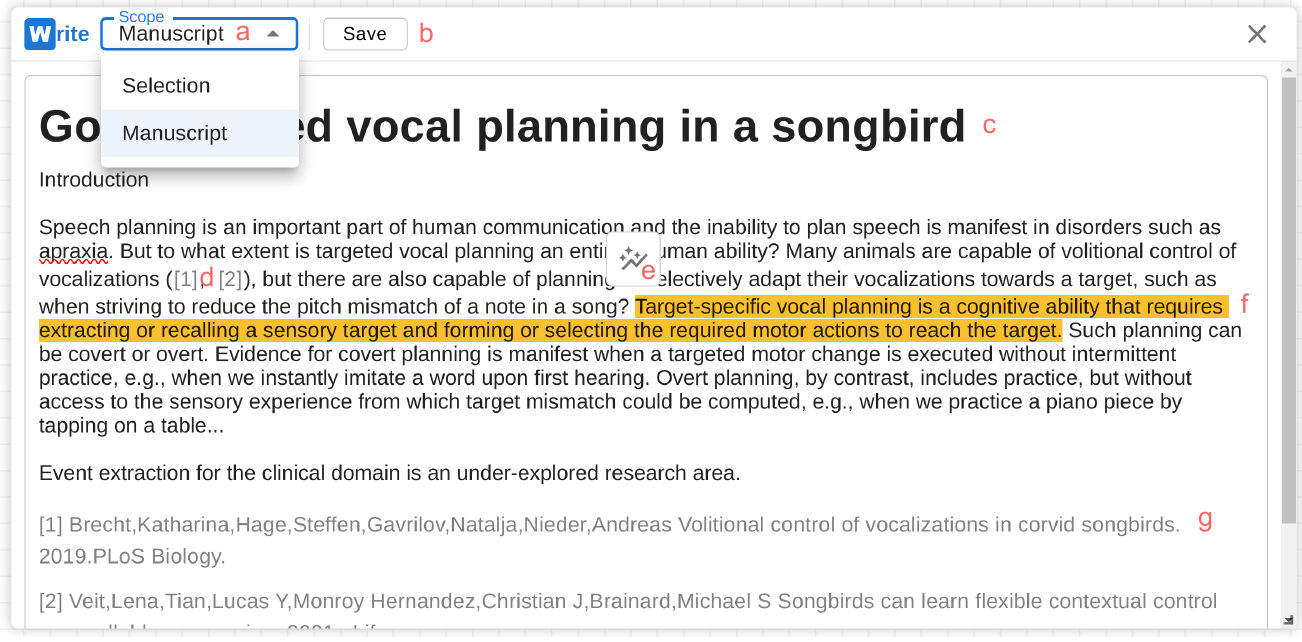}
    \caption{Overview of the Write module. (a) Scope selection for semantic search: opt for \texttt{Selection} to use user-selected text (highlighted in gold color) from the manuscript as the query for literature discovery, or choose \texttt{Manuscript} to search using the entire manuscript; (b) Manuscript saving: save the current version of the manuscript to the database; (c) Editable manuscript title: users can freely modify the title of their manuscript; (d) Citation marker: inserted from the document card in the Discovery module to reference relevant literature; (e) Highlight button: click to highlight the selected text; (f) User-selected text: highlighted in gold for visibility. To select text, click and drag the mouse, releasing at the end of the desired position; (g) Reference entries: displayed for each citation used within the manuscript. }
    \label{fig:write-module}
\end{figure}

\paragraph{Write (Figure~\ref{fig:write-module})} The Write module features an integrated text editor designed to streamline the research manuscript creation process. This module not only facilitates text composition but also enables users to conduct a semantic search using any selected content within the manuscript. This function efficiently retrieves the most relevant documents from the database, allowing users to concentrate on their writing without the distraction of a tedious document search process.

The Write module primarily functions as a hub for both retrieval and generation activities:

\begin{itemize}
    \item Retrieval: it utilizes the manuscript's content to conduct a semantic search, identifying highly relevant documents, or to pinpoint relevant content in another document, such as one in the Read module.
    \item Generation: the content of the manuscript is employed as part of the inputs for generating specific sentences, such as those for citations or arguments.
\end{itemize}

Users are provided with complete flexibility to select any portion of text within their manuscript as a basis for performing retrieval or generation functions as needed. This means they can choose specific sentences, paragraphs, or even entire sections of their document to initiate these tasks. Whether it's for extracting relevant information from other sources or generating new content such as citations or arguments, users have the freedom to use any part of their manuscript as input. This level of control ensures that the functions align closely with the users' current focus and the specific requirements of their work, enhancing the overall efficiency and relevance of the tasks performed.

\begin{figure*}[!htb]
    \centering
    \includegraphics[width=\textwidth]{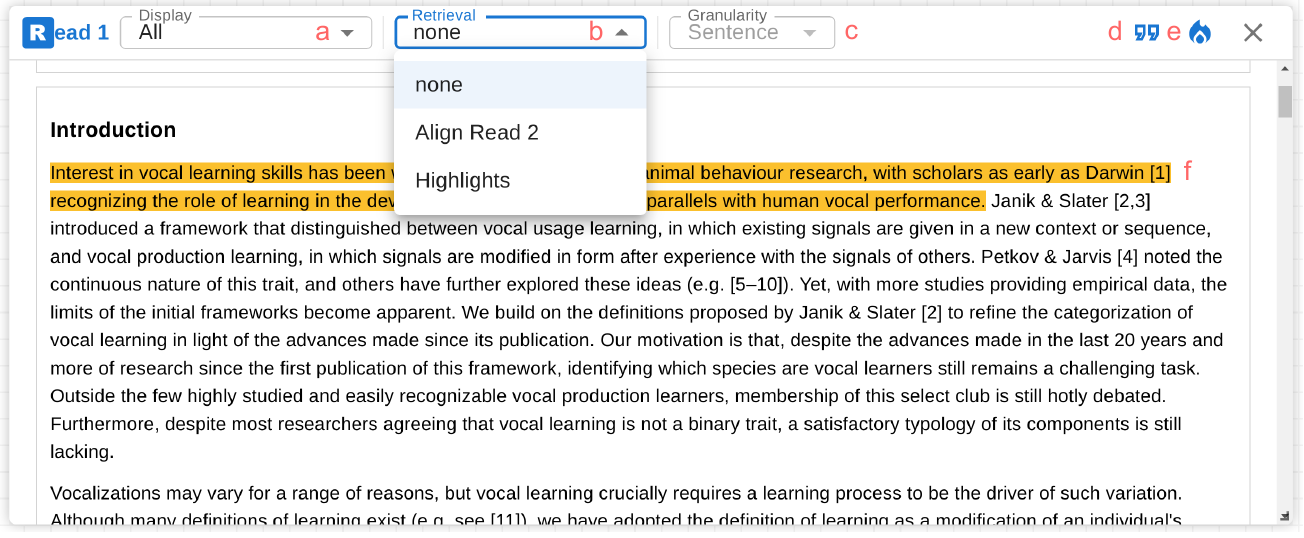}
    \caption{Overview of the Read module. (a) Section or document display option: users can choose to view either a specific section or the entire document; (b) Retrieval direction settings: selecting \texttt{none} allows the user to freely choose text from the current document for use as a text alignment source, or choose another document (e.g. \texttt{Align Read 2}) to set it as the text alignment target. The \texttt{Highlights} option provides a concise summary of the document being read; (c) Text alignment granularity: set the level of detail for text alignment results. The default setting is \texttt{Sentences}, with other available options being \texttt{Paragraphs} and \texttt{Sections}; (d) Cite button: click to cite this document in user's manuscript (citation marker added to the cursor position in Write module); (e) Local ``Fire'' button\fire:: activate the text alignment retrieval function; (f) User-selected text for alignment: this highlighting displays the text chosen by the user to serve as the source for text alignment.}
    \label{fig:read-module-source}
\end{figure*}

\paragraph{Read (Figure~\ref{fig:read-module-source})} A module that presents either the entire document or a specific section of it. It also highlights text at different levels by allowing authors to freely choose text for use in text retrieval and text generation functions.

\begin{figure*}[!htb]
    \centering
    \includegraphics[width=\textwidth]{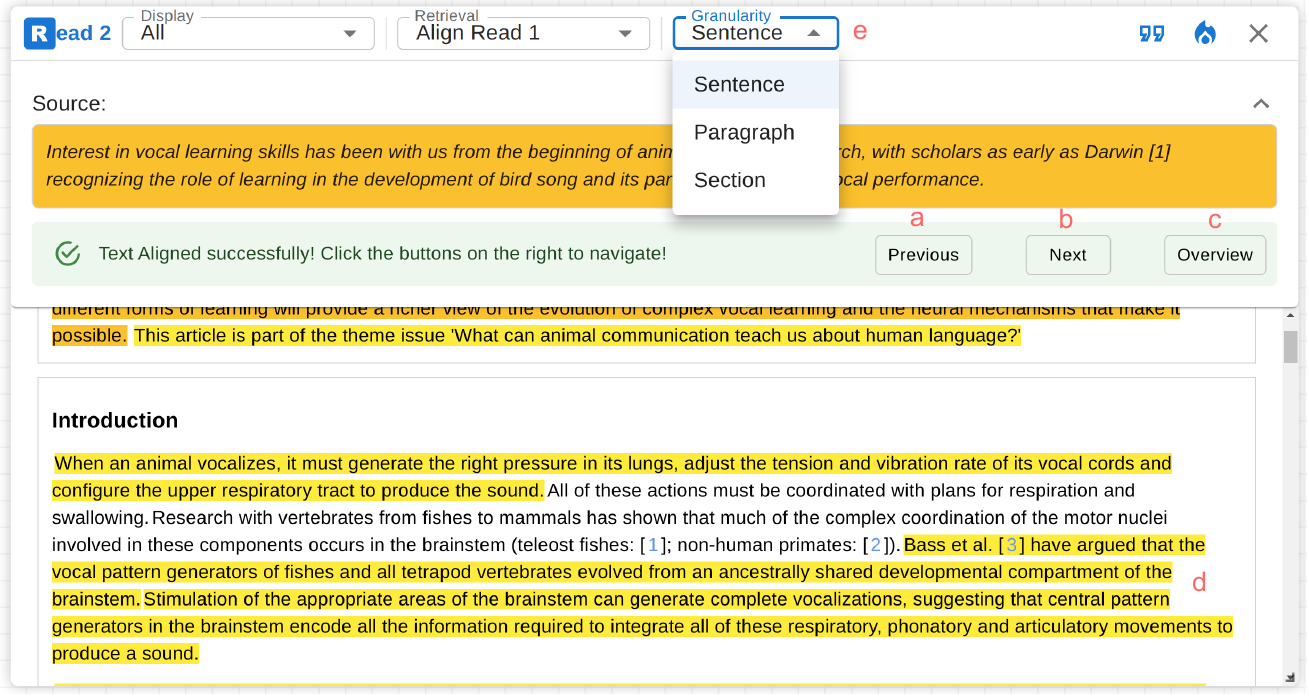}
    \caption{(a) and (b) navigation buttons: These buttons automatically scroll the window to the location of the aligned text within the document; (c) List view button: Displays all aligned text simultaneously for easy overview and comparison; (d) Aligned text display: Showcases text from the document that aligns with the selected text in the source document, as determined by the text alignment retrieval function. The brightness of the background color indicates the textual relevance; (e) Text alignment granularity: set granularity for the retrieved most relevant content.}
    \label{fig:read-module-target}
\end{figure*}

\begin{figure*}[!htb]
    \centering
    \includegraphics[width=\textwidth]{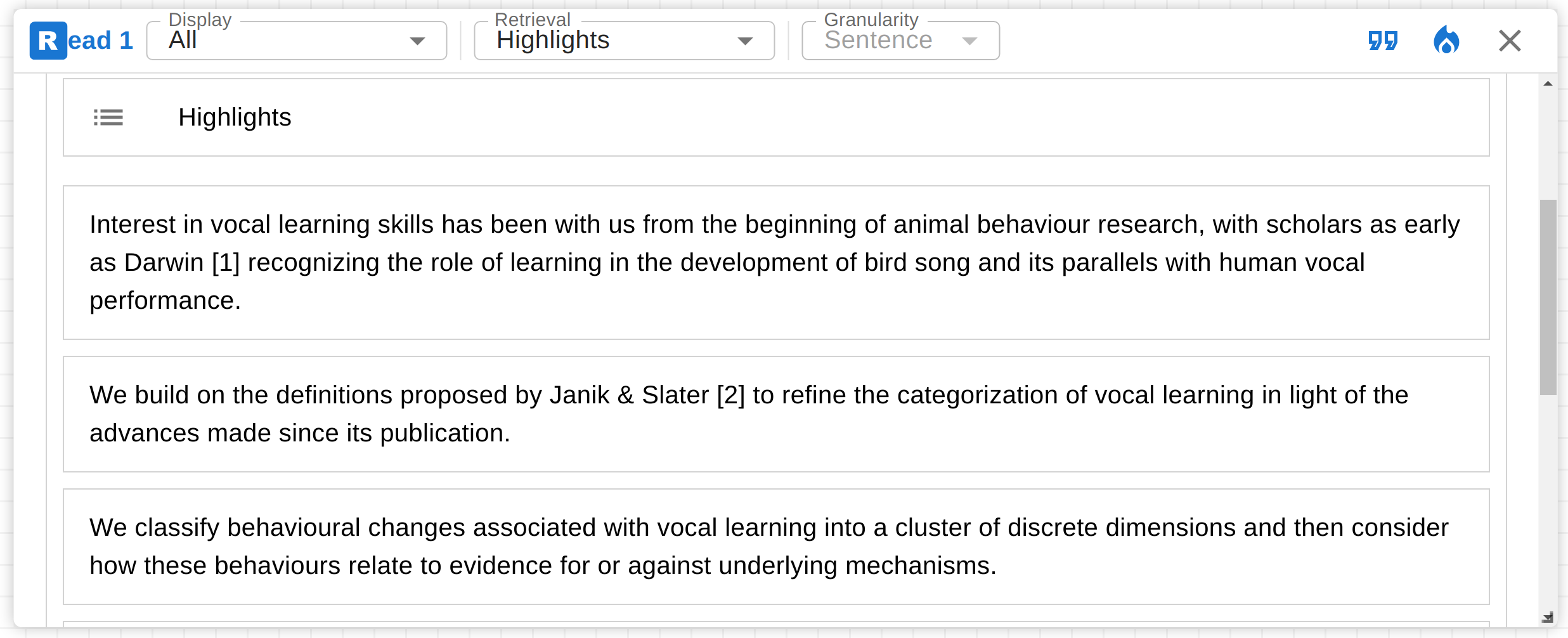}
    \caption{Read module for viewing the document summary. The summary is generated by an extractive text summarizer trained on scientific papers \cite{gu2022memsum}. Each entry is a complete sentence extracted from the body text of the current document. }
    \label{fig:read-module-summarization}
\end{figure*}

In the header of a Read module, there are three drop-down menus for setting up the viewing and retrieval functions:

\begin{itemize}
    \item Display: choose whether to view the complete document content or a specific section.
    \item Retrieval: configure the source and target document for text alignment function, or display the summary of the current viewing document. 
    \begin{itemize}
        \item To designate the current document as the source for text alignment, opt for \texttt{none} to allow for free text selection with the mouse within the current document. The selected text within the alignment source will be highlighted in a golden color.
        \item To set the current document as the alignment target, select the window ID of the alignment source (i.e. the Read module whose Retrieval menu is set to \texttt{none}, for example, \texttt{Align Read 2}). The text selected in the alignment source document will appear in the banner of the alignment target document, which authors can collapse by clicking the arrow icon. 
        \item To trigger the text alignment function, click on the Fire button\fire in the header. After retrieving the aligned text, authors can navigate through the aligned text using the PREVIOUS and NEXT buttons (see Figure~\ref{fig:read-module-target}).
        \item To access the summary of the current document, click on \texttt{Highlights}. Aligned texts will be displayed individually (see Figure~\ref{fig:read-module-summarization}). 
    \end{itemize}
    \item Granularity: establish the level for aligned text. Options include \texttt{Sentence}, \texttt{Paragraph}, and \texttt{Section}. Aligned texts will also be marked with a gold color (See Figure~\ref{fig:read-module-target}).
\end{itemize}

The Read module enables authors to flexibly define text as input to retrieval functions. It remains attentive to any alterations in the text chosen by the author. As depicted in Figure~\ref{fig:check-cite}, text alignment can occur between two instances of the Read module, a Write module, and a Read module, as well as a Generation module and a Read module.

\begin{figure*}
    \centering
    \includegraphics[width=\textwidth]{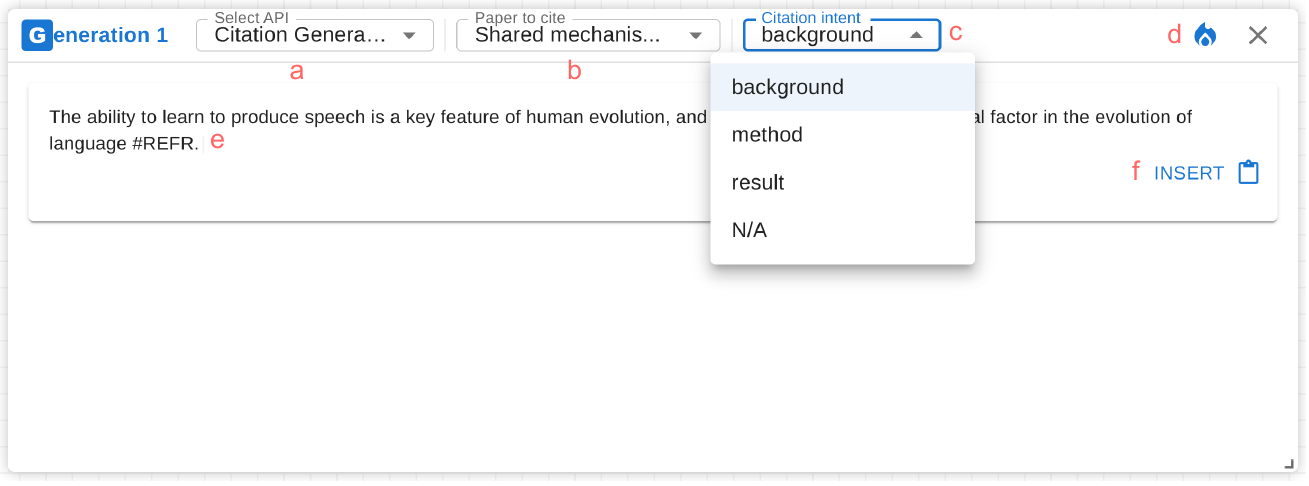}
    \caption{Overview of the Generation module (citation generation). (a) API selection menu: choose the specific generation API to be executed; (b) Citation generation API: select the document for which a citation sentence needs to be generated; (c) Citation intent selection: choose the intended purpose for the citation sentence being generated; (d) Local ``Fire'' button\fire:: activate the generation function to create citation sentences; (e) Display of generated citation sentence: the citation sentence appears with the citation marker denoted by \#REFR. (f) INSERT button: easily copy the generated citation sentence into the user's manuscript.}
    \label{fig:generation-module-citation}
\end{figure*}

\paragraph{Generation (Figure~\ref{fig:generation-module-citation})} This module seamlessly integrates text generation functions, leveraging text selections made by users in other modules. It is adept at producing diverse scientific texts, including citation sentences and argumentative sentences, thereby facilitating and enhancing the writing process. Additionally, the module functions as a plug-and-play hub for various text generation applications, providing a platform where cutting-edge generative models can be efficiently deployed.

\begin{figure*}
    \centering
    \includegraphics[width=\textwidth]{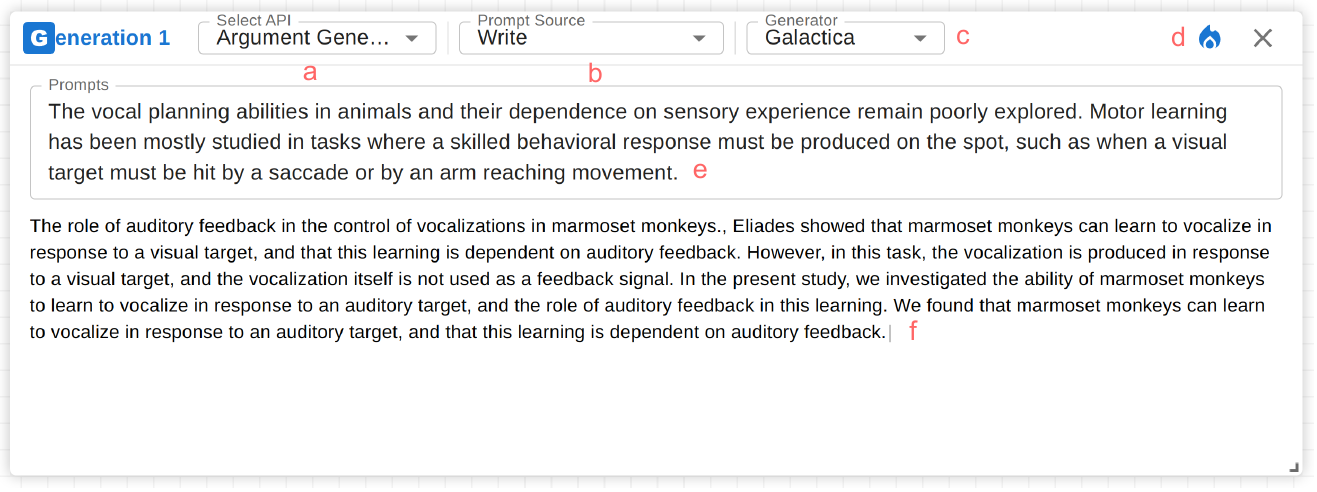}
    \caption{Overview of the Generation module (argument generation). (a) API selection menu: choose the specific generation API to be executed; (b) Argument generation API: choose the source from which the user-selected text will serve as the premise prompt for generating argumentative text; (c) Generative model selection: pick the specific generative model to use for creating the argumentative text; (d) Local ``Fire'' button\fire:: trigger the generation process for argumentative text; (e) Premise prompt input: displays the premise prompt used for the argument generation. Users have the flexibility to modify this prompt in the input bar; (f) Output of generated argumentative text: shows the text produced by the argument generation process. }
    \label{fig:generation-module-argument}
\end{figure*}
\begin{figure*}
    \centering
    \includegraphics[width=\textwidth]{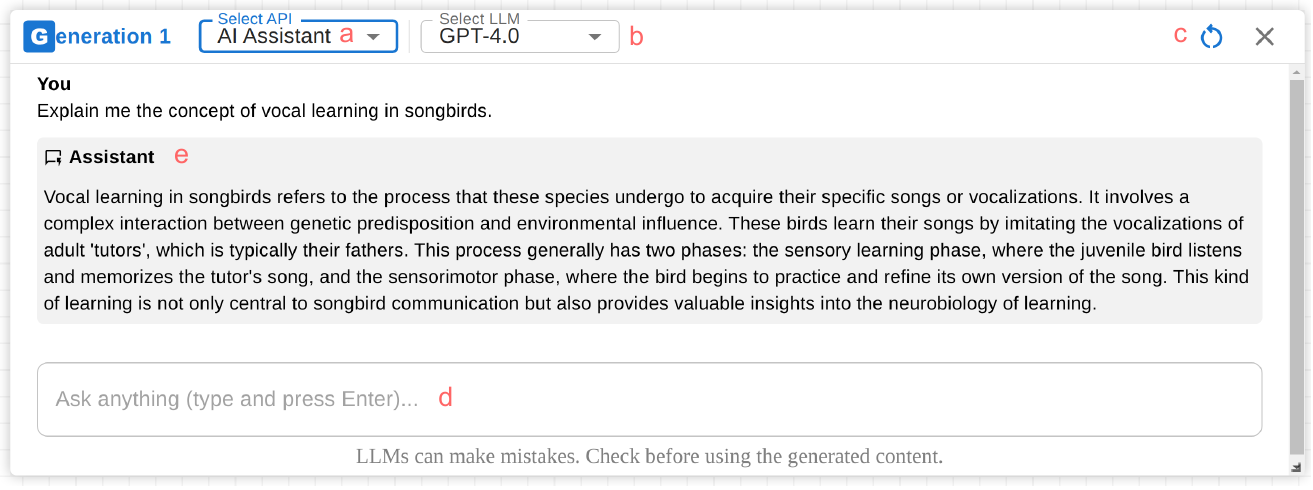}
    \caption{Overview of the Generation module (AI Assistant). (a) API selection menu: choose the specific generation API to be executed; (b) AI Assistant API: select the Large Language Model (LLM) to be used. We have integrated GPT-4.0 on the backend with self-designed prompts restricted to scientific writings; (c) Clear button: click to clear the conversation history; (d) Input bar for interacting: users can type in their query to interact with the AI assistant; (e) Display of AI assistant response: text generated by the AI assistant is displayed.}
    \label{fig:generation-module-assist}
\end{figure*}

Within the Generation module, we have incorporated two sophisticated text generation APIs:

\begin{itemize}
    \item Citation generation (\citet{gu2022controllable}, Figure~\ref{fig:generation-module-citation}): this API utilizes inputs such as keywords (from the Keywords module), contextual information (from the Read module), and a user-selected citation intent. It processes these inputs to generate a well-structured citation sentence.
    \item Argument generation (\citet{gao2024evaluating}, Figure~\ref{fig:generation-module-argument}): this API operates by taking a premise text selected by the user and then creatively generating a corresponding conclusion text, effectively forming a coherent argument.
    \item AI Assistant (Figure~\ref{fig:generation-module-assist}): this API allows users to interact with an LLM for general purposes. We have integrated GPT-4.0 \citep{achiam2023gpt} for beta testing and will bring more specialized LLMs into production.
\end{itemize}

The integration of these advanced APIs within the Generation module significantly enhances the user's research and writing capabilities. The citation generation API, particularly, streamlines the process of referencing, ensuring that citations are contextually relevant and accurately formatted. It not only saves time but also enriches the manuscript with authoritative sources. On the other hand, the argument generation API, offers a unique tool for developing persuasive and logically sound arguments. By generating conclusion texts based on the user's selected premises, aids in constructing robust and compelling narratives. This feature is especially beneficial for drafting complex argumentative essays or research discussions, where substantiating a point of view with coherent reasoning is crucial. 


It is essential to emphasize the importance of ethical practices when utilizing the output from the generative models. Users are strongly encouraged to thoroughly review and verify the generated content before incorporating it into their manuscripts. This step is crucial for maintaining the integrity and accuracy of their work. The review process ensures that the content aligns with factual information, adheres to relevant guidelines, and meets academic standards. By doing so, users can confidently use these advanced tools, while upholding ethical standards and contributing to the production of high-quality, reliable, and scholarly work.

\end{document}